Diffraction controlled backscattering threshold and application to Raman gap

Harvey A. Rose[1, 2] and Philippe Mounaix[3]

Abstract

The range of stimulated Raman scattering (SRS) frequencies covers a domain which at the low end abuts $\omega_{\min} = \omega_0/2$, according to the simplest SRS theories, corresponding to scatter from electron densities near ¼ critical. Experiments, on the other hand, clearly point to $\omega_{\min} > \omega_0/2$: SRS is *not* observed in a frequency gap between $\omega_0/2$ and $\omega_{\min}$, indicating a drastic disruption of scatter from Langmuir waves as electron densities approaches ¼ critical from below. Several one-dimensional mechanisms, linear and nonlinear, have been proposed to explain this "Raman gap". In this paper we release the one-dimensional constraint by allowing diffraction of the scattered light. In the linear convective regime we find that diffractive effects on SRS from a wide speckled laser beam tend to increase the SRS threshold with increase of density, so long as the interaction length is comparable to or larger than a speckle length. This may lead to a new, diffraction controlled, contribution to the Raman gap.

I.  **INTRODUCTION**

Ever since the theory[1] and observation[2] of stimulated Raman scattering (SRS) were enunciated, experimental SRS data[3, 4, 5] from plasma with peak electron density above $n_c/4$, with $n_c$ the laser critical electron density, often present a "gap" in the spectrum of scattered light, corresponding to an absence of SRS from densities close to $n_c/4$. In this paper we will show that diffraction of scattered light from a speckled laser beam, such as produced by a random phase plate[6] (RPP), leads to a qualitative increase of the SRS convective threshold gain exponent, $G_{0T}$ if the SRS interaction length is not small compared to a scattered light speckle length. Since diffractive effects increase with density, this tends to suppress SRS as density increases. However, since actual experimental conditions are so varied, and since exceedingly small levels of SRS reflectivity may be adequate to induce nonlinear SRS effects[7], diffraction, though universally applicable, cannot be the physically appropriate explanation of all gap manifestations. So far, diffractive effects on convective instability threshold have largely been ignored in standard instability analysis [8].

[1] New Mexico Consortium, Los Alamos, New Mexico 87544, USA
e-mail: hrose@newmexicoconsortium.org
[2] Theoretical Division, Los Alamos National Laboratory, Los Alamos, New Mexico 87544, USA
[3] Centre de Physique Théorique, UMR 7644 du CNRS, Ecole Polytechnique, 91128 Palaiseau Cedex, France
e-mail: mounaix@cpht.polytechnique.fr

The basic SRS model in the strongly damped convective regime is reviewed in section II. Calculational tools for determination of $G_{0T}$ in an RPP beam, and connections with critical intensity theory[9], are presented in section III. The dependence of $G_{0T}$ on plasma density is presented in section IV, immediately leading to a Raman gap, discussed in section V.

## II.  CONVECTIVE SRS WITH RPP OPTIC

The steady state laser light's electric field is assumed to propagate linearly in the positive "$z$" direction, in homogeneous plasma, with wavenumber $k_0$. Collisional absorption is ignored. In the paraxial wave approximation its spatial and temporal envelope, $E$, satisfies

$$i\frac{\partial E}{\partial z} = \frac{1}{2k_0}\Delta E .\qquad(1)$$

$\Delta$ is the transverse Laplacian, $\Delta = \partial^2/\partial x^2 + \partial^2/\partial y^2$, with periodic boundary conditions. $E$ has random phase boundary conditions, specified later. The scattered light with wavenumber $k_{SRS}$ and envelope $A$, satisifies[10]

$$i\frac{\partial A}{\partial z} = \frac{1}{2k_{SRS}}\Delta A + i\kappa|E|^2 A ,\qquad(2)$$

in the strongly damped Langmuir wave response regime. The spatial average of the laser intensity along the transverse directions, $I = |E|^2$, is normalized to unity so that $\kappa$, a constant, is the mean amplitude spatial gain rate.

### A. Inhomogeneous *interludium*

If fluctuations in plasma density, $\delta n$, vary linearly with position,

$$\delta n/n = z/L_n, \tag{3}$$

then laser light, scattered light and the SRS daughter Langmuir wave with damping rate $\nu_{Landau}$ and electron ion collision rate, $\nu_{ei}$, effectively stay in resonance over a distance $L_z$,

$$L_z \propto L_n \left(\nu_{Landau} + \nu_{ei}/2\right)/\omega_p. \tag{4}$$

If intensity fluctuations are ignored then $G_0 \propto 2\kappa L_z$, where $G_0$ is the resultant SRS power gain exponent. However, over the possibly large distance between low and ¼ critical density in real plasmas, the constant gradient model, Eq. (3), may fail. Therefore, in lieu of a detailed specification of the density profile (which in actual experiments is highly variable and typically not measured), we choose a slab model with independent parameters $G_0$ and system length, $L_z$. The local gain rate is not viewed as tied to local plasma properties but instead is given by

$$\kappa = G_0/2L_z. \tag{5}$$

Since, as we will show in section IV.C, Fig. 7, the larger the interaction length the stronger the effect of scattered light diffraction on the SRS threshold, diffraction will more strongly control the Raman gap the larger the ionic charge, at given values of other plasma parameters.

### B. RPP boundary condition

An RPP optic is modeled by an idealized boundary condition for *E* that has approximately uniform amplitude fluctuations for transverse wavenumbers $k < k_0/2F$ and negligible fluctuations for $k \gg k_0/F$, with *F* the optic f/#. Different Fourier modes have independent random phases. Such a spectrum and Eq. (1) then imply an electric field auto-correlation function, $C$, that on axis satisfies

$$C(z-z') \equiv \langle E(z,x)E^*(z',x) \rangle \sim 1/(z-z')^{(D-1)/2}, \qquad (6)$$

for $|z-z'| \gg F^2/k_0$, where $D$ is the space dimension. $C$ is normalized to unity at zero spatial separation. Since laser intensity, $I$, in the neighborhood of an intense speckle located at the origin of coordinates, varies as[11, 12]

$$I(x,z)/I(0,0) = |C(x,z)|^2, \qquad (7)$$

a speckle's gain is insensitive to $L_z$ once $L_z > L_{speckle}$, with speckle length defined by

$$L_{speckle} \equiv \int_{-\infty}^{\infty} |C(z)|^2 dz, \qquad (8)$$

finite and $\propto F^2/k_0$ for $D > 2$. Finite $L_{speckle}$ is important to preserve in 2D models, which we use for computational convenience, otherwise there would be an unphysical increase of SRS with $L_z$. Our specific 2D model RPP spectrum, with "$x$" the unique transverse direction, differs from the usual flat spectrum as $k \to 0$,

$$|E(k)| \propto \sqrt{k} \exp(-a^2 k^2), \qquad (9)$$

so that its $C(z)$ approaches zero at large $z$ as $1/|z|$, with $L_{speckle} = 4\pi a^2$. Physically, $a$ is a speckle radius, $a \propto F/k_0$. The precise proportionality coefficient between $a$ and $F/k_0$ is irrelevant for purposes of this paper because within the scope of our model, Eqs. (1) and (2), SRS gain only depends upon three dimensionless physical parameters: $G_0$, $L_z/L_{speckle}$ and $k_0/k_{SRS}$. Instead of separate notations for $E$ and its Fourier space representation, $x$ and $z$ arguments imply the former while $k$, $p$, and $q$ arguments imply the latter.

Results also depend implicitly on details of $E$'s boundary condition. A real RPP optic has a finite number of elements, $N_{RPP}$, that is proportional to the number of close packed speckles that fit into the laser beam's cross sectional area at best focus. For a 3D model with periodic boundary conditions, $N_{RPP}$ qualitatively corresponds to the number of Fourier modes with $k < k_0/2F$. In 2D, the correspondence is with the number of speckles encountered across the beam waist, which is proportional to the transverse periodicity length, $L_x$, scaled to a speckle's full width at half max, FWHM[13].

### III. THE SRS THRESHOLD DEFINITION (QUALITATIVE FORMULATION)

Strictly speaking, our model does not have an instability threshold because Eq. (2) predicts amplification of $A$ for any finite $\kappa$. One physically relevant definition of threshold requires that thermal fluctuations amplify to a finite level. For example, a nominal thermal reflectivity level[14], $R_0 = 10^{-8}$, would thus require $G_0 = \ln(10^8) \approx 18$ to attain reflectivity of order unity. However, this estimate ignores laser intensity fluctuations induced by the RPP optic, which can give rise to a significant overestimation of the threshold: 3D simulations suggest[15] that typical amplification is actually much larger than $\exp(G_0)$ once

$$G_0 \approx \max\left(1, L_z/L_{\text{speckle}}\right) \tag{10}$$

But once $L_z/L_{\text{speckle}} > \ln R_0^{-1}$, then intensity fluctuations are of diminished physical significance, *e.g.*, if $R_0 = 10^{-8}$ and $L_z/L_{\text{speckle}} > 18$ then the gain obtained without intensity fluctuations is sufficient to get a reflectivity of order unity. For the same reason, a similar conclusion would be reached if the threshold was alternatively defined as the average gain for the onset of nonlinear effects.

Instead of threshold based on attaining a certain amplification, we adopt a definition of the threshold gain exponent, $G_{0T}$, that compares the contribution of the higher laser intensity fluctuations to SRS reflectivity, *R*, with the contribution of the complementary lower fluctuations: above (below) threshold the former (latter) dominate. A quantitative practical formulation will be given in the next section. This definition may be used to identify a physical limit to the linear model's validity: below threshold, one expects that linear estimates are not qualitatively affected by nonlinear effects in rare, very intense fluctuations, and *vice versa*. Since rare events have large fluctuations, *R* is expected to have large fluctuations[16] near threshold. It is worth mentioning that our threshold definition is free of any hot-spot model approximation. This is an important point as such a model may be questionable[17] for long systems.

### A. Threshold gain versus critical gain

Let $\langle G \rangle$ be the mean power gain exponent,

$$\exp\langle G \rangle = \left\langle \int |A(x,z=L_z)|^2 \, dx \right\rangle \bigg/ \int |A(x,z=0)|^2 \, dx = \langle R \rangle / R_0 \tag{11}$$

The incident power, $R_0 \propto \int |A(x,z=0)|^2 \, dx$, is assumed to have negligible fluctuations and the expectation value, $\langle \ \rangle$, denotes the average over different laser boundary realizations. If the $\{E(k,z=0)\}$ are independent, Gaussianly distributed, complex random variables with zero mean and mean square value given by, e.g., Eq. (9), then it has been rigorously shown that for given $L_z$ there is a critical value of $G_0$, $G_{0c}$, and a corresponding critical value of $\kappa$, $\kappa_c$, such that:

1. $\langle G \rangle$ diverges[9] for $\kappa > \kappa_c$
2. $\kappa_c L_{\text{speckle}}$ is independent[9] of $k_0/k_{\text{SRS}}$
3. $\kappa_c$ is a non-increasing[18] function of spatial dimension for certain choices of $C(x,z)$.

Though the first consequence of the Gaussian model cannot be interpreted literally with regard to real experiments because basic principles limit $R$ to finite values, the model's rapid increase of $R$ as $\kappa$ increases past $\kappa_c$ may be physically relevant. However, the last two consequences are not born out by simulations of an experimentally more appropriate RPP model, one in which Fourier mode amplitudes are fixed, while their phases are random. This will simply be referred to as the "RPP model". Expectations based on the central limit theorem that these models have similar properties for wide enough laser beams are tempered by the fact[19] that the RPP model's power amplification may become non-ergodic as $\kappa$ increases past $\kappa_c$ for any beam width. It is not the purpose of this paper to show if and how the Gaussian and RPP models' results can be reconciled. Instead our alternative definition of threshold may be viewed as a generalization of the independent hot spot model definition. In section IV this definition is quantified and applied to RPP simulation results.

### B. How diffraction is expected to affect $G_{0T}$

First, consider the idealized Gaussian model in which $G_{0T}$ is equated with $G_{0c}$. In this case it has been proved[9] that for any finite non-zero $k_0/k_{SRS}$ and fixed $F^2/k_0$, $G_{0T}$ is a constant of $k_0/k_{SRS}$ (see Property 2 of $\kappa_c$ above) which cannot be greater than the critical gain exponent without diffraction, at $k_0/k_{SRS} = 0$. Technically, the latter is given by

$$\kappa_c(L_z) = 0.5/\mu_{max}(L_z) \tag{12}$$

where $\mu_{max}$ is the largest eigenvalue, of $C(z-z')$ with $0 \leq z, z' \leq L_z$, i.e., the largest $\mu$ such that

$$\int_0^{L_z} C(z-z')\phi(z')dz' = \mu\phi(z).$$

admits a square integrable solution. On the other hand, for a density arbitrarily close to $n_c/4$, $k_0/k_{SRS}$ gets arbitrarily large, and in this limit it can be shown that $G$ reduces to $(2\kappa/S)\int |E(x,z)|^2 dx\,dz$, where $S$ is the plasma cross section (width in 2D, area in 3D). The corresponding threshold gain exponent is then found to be given by Eqs. (5) and (12) where $\mu_{max}$ is now the largest eigenvalue of $S^{-1}C(x-x',z-z')$, with $0 \leq z, z' \leq L_z$, i.e., the largest $\mu$ such that

$$\int dx' \int_0^{L_z} C(x-x',z-z')\phi(x',z')dz' = S\mu\phi(x,z)$$

admits a square integrable solution. Now, it turns out that the largest eigenvalue of $C(z-z')$ cannot be less than the largest eigenvalue of $S^{-1}C(x-x',z-z')$ and one gets

$$G_{0T}(n/n_c = 1/4) \geq G_{0T}(0 \leq n/n_c < 1/4) = G_{0T}(n/n_c = 0) \tag{13}$$

where $n/n_c = 0$ (resp. ¼) corresponds to $k_0/k_{SRS} = 1$ (resp. $+\infty$). Figure 1 (red line) shows the typical behavior of $G_{0T}$ as given by Eq. (13). The displayed results can be explained simply as follows: in the Gaussian model $G_{0T}$ corresponds to the divergence of $\langle G \rangle$ which is determined by arbitrarily high laser intensity fluctuations. So, whatever $k_0/k_{SRS} \geq 1$, finite diffraction effects are not strong enough to affect the value of $G$ in such extreme fluctuations significantly, accounting for the constant $G_{0T}$. It is only for $k_0/k_{SRS} = +\infty$, at $n = n_c/4$, that diffraction can compete against amplification in arbitrarily high intensity fluctuations, thereby increasing $G_{0T}$.

The situation is different when a cutoff limiting the largest possible $G$ is imposed on the Gaussian model. Since laser intensity fluctuations are now bounded, finite diffraction effects can affect the value of $G$ even in the highest fluctuation which is now *finite*. And the larger $k_0/k_{SRS}$ the more important the effect. Thus, diffraction is expected to smooth the discontinuity of the Gaussian model threshold at $n = n_c/4$, turning it into a monotonically increasing boundary layer near $n = n_c/4$. This behavior is shown qualitatively in Fig. 1 (green solid line). The slight decrease of $G_{0T}$ from Gaussian model $G_{0c}$, the red dot in Fig. 1, at $n = n_c/4$, is to be expected from the asymptotic form of the distribution of $G$ at large, but finite, values. (This effect diminishes when the cutoff value increases).

Consider now the RPP model with the same $C(x,z)$. Since in this model the spatial average of the laser intensity (along the transverse directions) is nonrandom and normalized to unity, the expression of $G$ at $n = n_c/4$ reduces to the nonrandom finite value $2\kappa L_z = G_0$, whatever the number of modes. It follows that in the case of an infinite number of modes where laser intensity fluctuations are not bounded and the concept of critical intensity is still meaningful, one has $G_{0c}(n/n_c = 1/4) = +\infty$ and the red dot in Fig. 1, at $n = n_c/4$, goes to infinity. Note that the same result would be obtained in the Gaussian model with $S \to \infty$ (by ergodicity of the Gaussian field). Similarly, for a finite number of modes one expects $G_{0T}(n/n_c = 1/4) = +\infty$ and the curve of $G_{0T}$ as a function of $n/n_c$ must have a vertical asymptote at $n = n_c/4$ (see the green dotted line in Fig. 1).

As can be seen in Fig. 1, for any $G_0$ not too large there is a density interval near and below $n_c/4$ where $G_0$ is below threshold, giving rise to a diffraction controlled Raman gap. We recall, as discussed in section II.A, that this result is in the context of a slab model: as $n$ increases, $G_0$, is fixed.

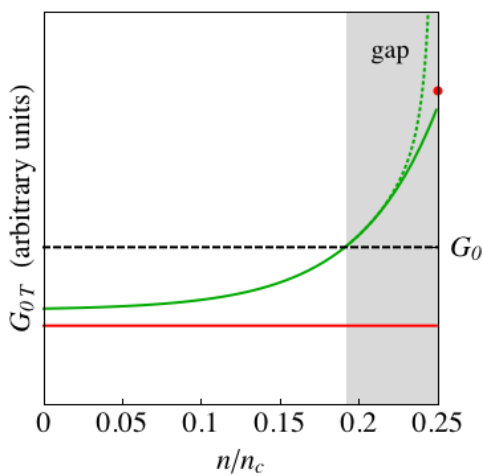

Fig. 1: Typical behavior of critical gain exponent in the Gaussian model (red) and expected typical behavior of threshold gain exponent in the Gaussian with cutoff model (green solid line) and in the RPP model (green dotted line) as functions of electron density. A given $G_0$ is below threshold in the gray region, which defines a diffraction controlled gap.

## IV. THE GAIN SPECTRUM FORMALISM

Since the scattering model, Eq. (2), is linear in $A$, the denominator in Eq. (11) may be replaced by unity. For a given realization of $E$'s phases at the boundary $z = 0$, and with a slight abuse of notation, the "reflectivity", $R$, is given by

$$R = \int |A(x,z = L_z)|^2 \, dx \qquad (14)$$

The Fourier components of $A$ at $z = L_z$, $a_j(L_z) \equiv A(\mathbf{k}_j, L_z)$, are linearly related to those at $z = 0$,

$$a_l(L_z) = \sum_{|k_j| < k_{max}} P_{lj} a_j(0), \qquad (15)$$

where the propagator matrix, $\mathbf{P}$, depends in detail on $|E(x,z)|$, through the solution of Eq. (2). Allowed boundary conditions are wavenumber limited because otherwise the paraxial wave equation loses its physical validity. This wavenumber limit, $k_{max}$, is an additional parameter of the model, typically chosen as $k_{max} = k_0/F$, though results presented later are not much affected by doubling this limit. The Fourier index, $l$, for the transmitted light, is not limited in principle. In practice, its limit is determined by the discretization in $x$, $dx$, of a simulation, $|k_l| < \pi/dx$. Eqs. (14) and (15) imply, using the Dirac[20] bra and ket notation, that

$$R = \langle a(0) | \mathbf{R} | a(0) \rangle \tag{16}$$

$$\mathbf{R} = \mathbf{P}^\dagger \mathbf{P} \tag{17}$$

The reflectivity matrix, $\mathbf{R}$, is Hermitian with real positive eigenvalues, $\{r_j\}$, conveniently represented by their gain coefficients, $g_j = \ln(r_j)$. For any RPP realization and given $g > 0$, there are a certain number of $g_j$ between zero and $g$. When averaged over RPP phases, this number defines a cumulative distribution for $g$, the (formal) derivative of which with respect to $g$ determines the mean gain coefficient spectral density, $\langle \rho(g) \rangle$. Its properties follow from the RPP's detailed statistics and, loosely speaking, all possible solutions of Eq. (2). Since the eigenvectors of $\mathbf{R}$ are orthogonal, it follows that with suitably normalized broadband noise as a boundary condition for $A$ at $z = 0$,

$$\langle R \rangle = \sum_j \langle \exp(g_j) \rangle = \int_0^\infty \langle \rho(g) \rangle \exp(g) dg, \tag{18}$$

with the bracket now signifying average over the realizations of both the RPP and $a(0)$

### A. Quantitative formulation of the SRS threshold

The integral representation, Eq. (18), allows a quantitative formulation of the SRS threshold definition, discussed at the beginning of Section III. Let

$$\langle R(g) \rangle = \int_0^g \langle \rho(s) \rangle \exp(s) ds \tag{19}$$

Consider a finite sample of RPP realizations and estimate $\langle \rho(g) \rangle$ from that sample. Let $g_{max}$ denote the largest value attained by $g$ in the sample. Since the sample is finite, $g_{max}$ is finite with probability one. The idea is to link the SRS threshold definition with a change in the behavior of the estimated $\langle R(g) \rangle$ at $g_{max}$. Note that in the limit of an infinite sample, $g_{max} \to \infty$ and the idea is consistent with the definition of the critical gain $G_{0c}$: if $\langle R(g_{max}) \rangle$ is convergent (divergent) as $g_{max} \to \infty$, then $G_0$ is below (above) $G_{0c}$. Of course, the simulation results presented later involve a finite sample and a practical alternative to the convergent/divergent criterion is required. First, we specify the way numerical data are processed. Bin the $\{g_j\}$ from every realization into equally spaced intervals, $[g_{min} + i \times dg, g_{min} + (i+1) \times dg]$, $i = 1, N$, with parameters $g_{min}$, $dg$ and $N$ chosen to include all realized values of $g_j$, $g_{min} \le g_j \le g_{max}$, for every $j$ and every realization. Let $g_{sig}$ be the largest value of $g$ (to within $dg$) such that the number of $g_j$ values in its bin is statistically significant. In practice we compromise at 10. Such a compromise is necessary because as an approximation to $\langle R \rangle$, $\langle R(g) \rangle$ should be evaluated for $g$ large as possible, at $g = g_{max}$, but the requirement of using a statistically significant estimate points to $\langle R(g_{sig}) \rangle$. We are now ready to give our definition of $G_{0T}$. As explained at the beginning of Section III, below threshold the reflectivity is determined by the contribution of the bulk of the $g_j \sim \langle g \rangle$, and since $g_{sig} > \langle g \rangle$, $\langle R(g) \rangle$ must be concave down at $g = g_{sig}$. Assuming that the reciprocal is true for the physical problem we consider, one is led to the following definition: if $\langle R(g_{sig}) \rangle$ is concave down (up) at $g = g_{sig}$, then $G_0$ is below (above) $G_{0T}$ (to be compared with the definition of $G_{0c}$ above).

### B. Relation to independent hot spot model

It is interesting to note that Eq. (18) keeps the same form as the expression of the reflectivity obtained from the much simpler independent hot spot (IHS) model[15]. In this model, reflectivity is assumed to be dominated by the contribution of independent intense speckles for which the gain is well approximated by (recall that the mean intensity has been normalized to unity)

$$g = 2\kappa L_{speckle} I_{speckle} = \langle G_{speckle} \rangle I_{speckle} \qquad (20)$$

where $\langle G_{speckle} \rangle$ is the average 1D gain over one speckle length. Aside from algebraic corrections[21, 22], the distribution of speckle intensities is asymptotically exponential, so that neglecting multiple amplification in rare high intensity speckles, one arrives at the independent hot spot (IHS) model result,

$$\langle R \rangle \sim \int_0^\infty \exp(-g/\langle G_{speckle} \rangle) \exp(g) dg. \tag{21}$$

This expression is similar to Eq. (18) with $\langle \rho(g) \rangle \sim \exp(-g/\langle G_{speckle} \rangle)$. In a finite system the highest speckle intensity is finite and typically of the order of the logarithm of the system size. This provides a natural order of magnitude for $g_{max}$. Now, it is easily seen that $\langle R(g_{max}) \rangle$ is concave down (up) if $\langle G_{speckle} \rangle < 1 (>1)$. Applying then our threshold definition we find that the IHS model approximation to $G_{0T}$ is simply given by the IHS model approximation to $G_{0c}$. For a long system, the latter can be significantly different from the exact $G_{0c}$ and the model may be quantified[17] to include the dependence on system length. Unfortunately, results are still at odds with the mathematics of $G_{0c}$. In addition, the apparent dependence of $G_{0T}$ on $k_0/k_{SRS}$ as revealed by simulation results presented later, is not captured by inclusion of diffractive corrections[23] to single speckle gain in the IHS. It should also be mentioned that improved IHS models lead to rather cumbersome, complicated, versions of Eq. (21). The gain spectrum formalism makes it possible to stick to the simple form Eq. (18). In this formal sense, it can be regarded as the proper generalization of the IHS model. The price to pay is the loss of a simple correspondence between the $g_j$ and some specific laser field structures (like intense speckles), making the interpretation of the results less intuitive.

### C. Small versus large gain rate regimes

The threshold algorithm is illustrated with two examples, both with $k_0/k_{SRS} = 1$. The SRS boundary condition wavenumber cutoff is $k_{max} = 2a$ and the transverse periodicity length, $L_x/FWHM \approx 140$. As a result, the dimension of the reflectivity matrix **R** is 201. The bin width is $dg = 0.073$ and the number of bins is 100.

The first example is in the weak gain rate, large length, regime, $G_0 = 5$ and $L_z/L_{speckle} = 40$. In this case, $\langle G_{speckle} \rangle = 1/8$ (see Eq. (20)), which is small compared to unity, the IHS model approximation to the threshold gain. One can thus reasonably expect SRS to be below threshold (in our sense) in this regime, which we will now check. Fig. 2 shows an estimation of $\langle \rho(g) \rangle$ obtained from 1,300 independent RPP realizations.

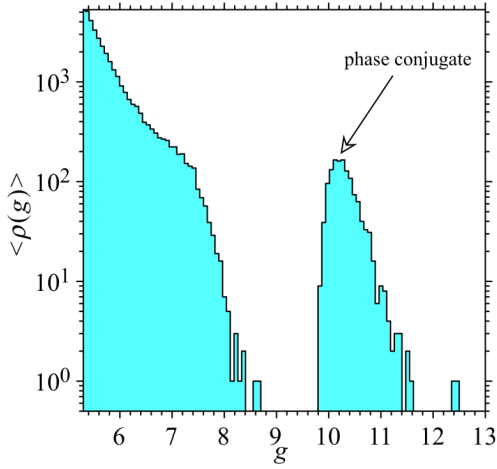 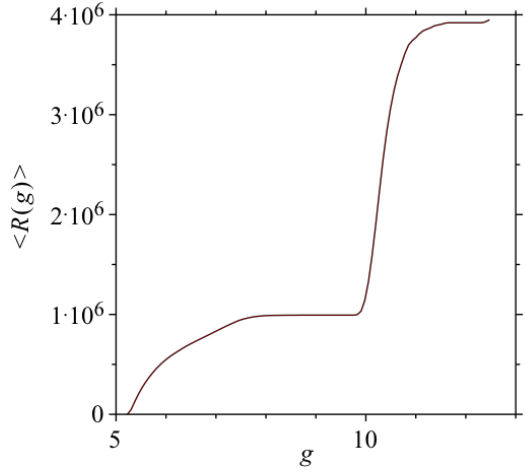

Fig. 2                 Fig. 3

Fig. 2. The gain spectral density for $k_0/k_{SRS} = 1$, $G_0 = 5$ and $L_z/L_{speckle} = 40$, manifests a peak near the phase conjugate gain value, $g = 2G_0$.

Fig. 3 caption: The cumulative reflectivity, $\langle R(g) \rangle$, is dominated by $\langle \rho(g) \rangle$'s phase conjugate peak. $\langle R(g) \rangle$ is clearly concave down at $g \approx 11$, the largest $g$ such that the numerically determined $\langle \rho(g) \rangle$ is statistically significant.

Here, $\langle \rho(g) \rangle$ is not normalized to unity. Instead, in each gain bin, its value is the total number of gain coefficients in that gain interval, from all realizations. Note the spectral peak near the phase conjugate[24] value, $g = 2G_0$. Details of the peak near $g = G_0$ depend on the dimension of **R**, hence on the value of $k_{max}$: the larger $k_{max}$, the larger this peak. For too large a $k_{max}$, these details are of course an artifact of the paraxial wave approximation. This dependence of the low part of the spectrum on the wavenumber cutoff does not affect evaluation of the cumulative reflectivity's concavity[25] at large $g$. In this regime the latter is clearly concave down, as can be seen in Fig. 3, and SRS is unambiguously below threshold, as expected.

The second example is in the moderate gain rate regime, $\langle G_{speckle} \rangle = 3/2$, and various $L_z$. Since $\langle G_{speckle} \rangle$ is order unity, no simple assertion can be made *a priori* as to SRS being above or below threshold. This is where our threshold definition can come into its own. Figs. 4 and 5 show estimations of $\langle \rho(g) \rangle$ obtained from 2,000 independent RPP realizations at $L_z/L_{speckle} \approx 0.4$ and $L_z/L_{speckle} \approx 2.0$ respectively.

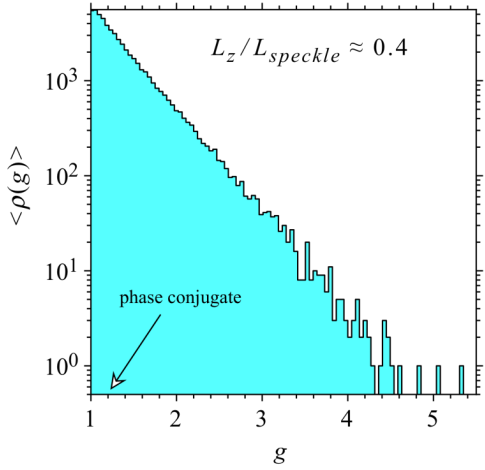 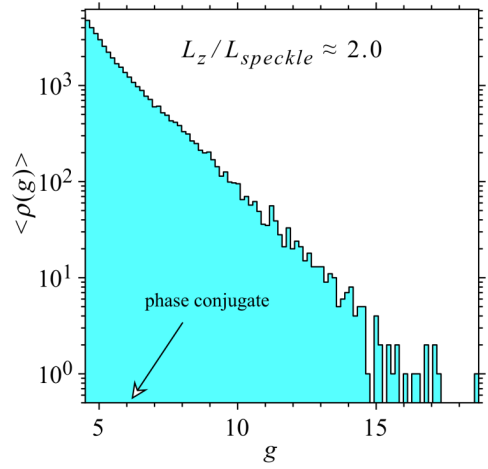

Fig. 4        Fig. 5

Fig. 4. The gain spectral density, for moderate gain rate and short propagation distance, $k_0/k_{SRS} = 1$, $\langle G_{speckle} \rangle = 3/2$ and $L_z/L_{speckle} \approx 0.4$, does not manifest a spectral feature at the phase conjugate mode value, indicated by the arrow.

Fig. 5. The gain spectral density for moderate gain rate, $\langle G_{speckle} \rangle = 3/2$, and a larger propagation distance, $L_z/L_{speckle} \approx 2.0$, (always with $k_0/k_{SRS} = 1$).

First, it can be seen that there is no specific spectral feature left near the phase conjugate value, $g = 2G_0$, indicated by an arrow in the figures: increasing the gain rate has completely destroyed phase conjugation. Now, while except for scale, these figures may appear similar, their cumulative reflectivity graphs are qualitatively different, as seen in Fig. 6. Each graph is normalized to unity at the largest statistically significant $g$ in its associated spectral density. For the parameters of Fig. 6, we judge that the threshold propagation distance, $L_T$, at which the curves change from concave down to concave up, is in the interval $1.2 < L_T/L_{speckle} < 1.6$. Recall that propagation of laser and scattered light is in 2D, and the laser spectrum is pseudo-3D, Eq. (9).

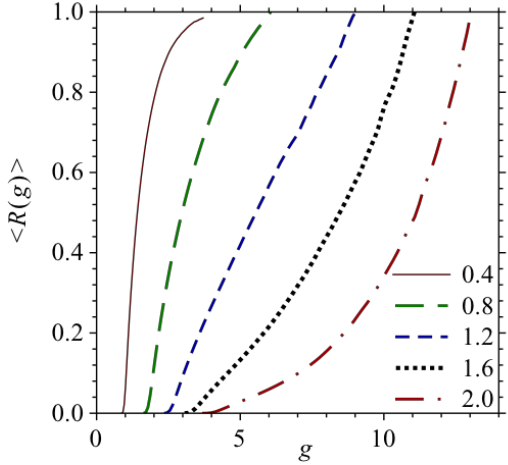 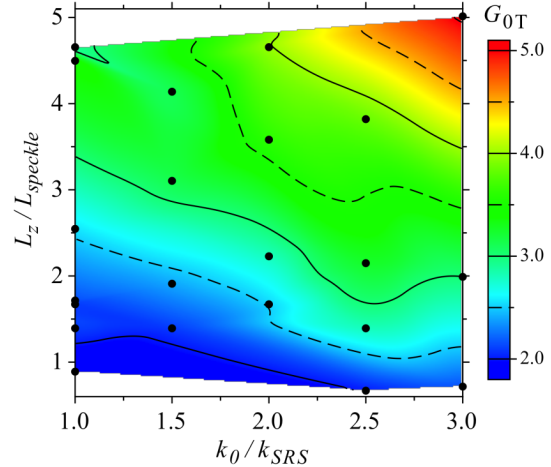

Fig. 6                                          Fig. 7

Fig. 6. The normalized cumulative reflectivity for different propagation distances, $L_z$, with $\langle G_{speckle} \rangle = 3/2$ and $k_0/k_{SRS} = 1$. Curves are labeled by $L_z/L_{speckle}$.

Fig. 7. SRS threshold gain exponent, $G_{0T}$, only depends upon the ratio of laser to scattered light wavenumbers and the slab thickness, $L_z$, scaled to the speckle length. Simulation data is represented by the black dots, while nearest neighbor interpolation was used to fill out the surface.

## V. THRESHOLD GAIN SIMULATIONS

Simulations were performed for various $L_z/L_{speckle}$ and $k_0/k_{SRS}$. In practice, a value of $\kappa$ is chosen and the cumulative reflectivity is evaluated at various propagation distances. As in the preceding section (see Fig. 6), an $L_T$ estimate is obtained from which the corresponding threshold 1D gain exponent, $G_{0T} = 2\kappa L_T$, follows. Errors in these estimates are both statistical and interpolative: fluctuations in any finite sample lead to uncertainties of $\langle R(g) \rangle$'s concavity at $g = g_{sig}$ (statistical errors), which combine with errors in $L_T$'s estimation owing to $\langle R(g_{sig}) \rangle$'s evaluation at finitely separated $L_z$ values (interpolative errors). For the results presented here, this translates into a 20% uncertainty of $G_{0T}$ which is mainly interpolative. Fig. 7 shows the estimated threshold gain exponent versus $k_0/k_{SRS}$ and scaled propagation distance. Note that a vertical cut through $k_0/k_{SRS} = 1$ also determines the backscatter SBS threshold. Now, the SRS backscatter frequency and wavevector matching conditions[8] determine $k_0/k_{SRS}$ as a function of $n/n_c$ and electron temperature, $T_e$. More precisely, we first solve

$$k_{SRS}\lambda_D = \sqrt{\frac{T_e}{511\,\text{keV}}} \sqrt{\left(\frac{1}{\sqrt{n/n_c}} - \frac{\omega_L(k)}{\omega_p}\right)^2 - 1}\,,$$

$$k\lambda_D = \sqrt{\frac{T_e}{511\,\text{keV}}} \sqrt{\frac{1}{n/n_c} - 1} + k_{SRS}\lambda_D\,,$$

numerically for $k$ and $k_{SRS}$ as functions of $n/n_c$ and $T_e$, where $\lambda_D$ denotes the Debye length and $\omega_L(k)$ is the angular frequency of the SRS daughter Langmuir wave (with wave vector $k$). The latter is deduced from the Langmuir wave kinetic dispersion relation[26]. The expression of $k_0/k_{SRS}(n/n_c, T_e)$ is then obtained readily from $k_0/k_{SRS} = k/k_{SRS} - 1$. This result, together with Fig. 7, determine $G_{0T}(n/n_c, L_z/L_{speckle}, T_e)$, whose graph is shown in Fig. 8 for $L_z/L_{speckle} = 4.5$ and two $T_e$ values: $T_e/\text{keV} = 1$ and $T_e/\text{keV} = 4$. This figure is similar to Fig. 1 (the green lines), which confirms the expected behavior of the threshold, as a function of $n/n_c$, explained in Sec. III B. In particular, it implies the possibility of a Raman gap, as shown in Fig. 1.

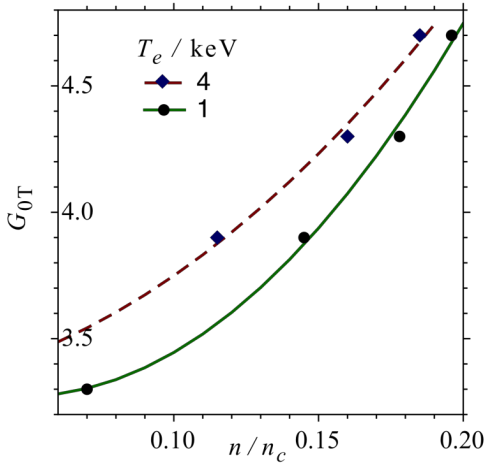

Fig. 8

Fig. 8. Increase of scattered light diffraction with density results in a 1D threshold gain exponent, $G_{0T}$, that rapidly increases with plasma density in the neighborhood of quarter critical.

It is in principle possible to go to higher $n/n_c$, $0.2 < n/n_c < 0.25$, but $k_0/k_{SRS}$ rapidly increases, as does the computational cost for determining $G_{0T}$, and the continuation of the graphs to larger density is left to the reader's imagination, guided by Fig. 1.

## VI. SUMMARY

We have determined the SRS backscatter gain exponent threshold, $G_{0T}$, in a speckled laser beam taking diffraction into account. Diffraction, always a player, has often been ignored because multi-dimensional simulations are more difficult than one dimensional, and because once laser speckles are allowed, huge reflectivity fluctuations are typical as the threshold is approached, making reflectivity based data analysis problematic. We have surmounted this reflectivity fluctuation issue by using a novel gain spectrum formalism in which the reflectivity takes a remarkably simple form, the integral (18). By considering the associated cumulative reflectivity (19), we have found that its concavity at the largest statistically significant gain value changes from concave down to concave up when the average 1D gain exponent $G_0$ increases past a certain value, $G_{0T}$, defining our SRS backscatter gain exponent threshold. By means of this definition, we have shown that diffraction alone causes the threshold to rapidly increase with plasma density in the neighborhood of quarter critical, implying a diffraction controlled "Raman gap". The larger the density gradient scale length, the larger is the gap. Since other mechanisms[27, 28, 29, 30, 31, 32, 33] may also contribute to a gap, detailed knowledge of plasma conditions is required to know if any one is dominant.

In our theory, the laser beam optic is modeled as an idealized random phase plate resulting in a speckled beam whose different Fourier modes have independent random phases but fixed amplitudes. Since the beam cross section is finite, speckle intensity fluctuations are bounded, and this model therefore yields a qualitatively different threshold behavior than does the critical gain theory of the Gaussian fluctuation model[9].

Long scale length experiments[34, 35] have observed a dependence of the stimulated Brillouin backscatter threshold on optic f/# that is in qualitative agreement with the independent hot spot model[15]. No comparable SRS threshold data is apparently available. Since the threshold depends on the optic f/#, or in other words, on the laser's coherence (speckle) length, any LPI that affects this coherence will obscure interpretation of an observed SRS threshold with the predictions of our theory. Therefore, sufficient laser bandwidth is needed[36] to suppress speckle self-focusing and sufficient ion-acoustic damping is required to suppress collective[37] forward stimulated Brillouin scatter (beam spray), both of which decrease the laser's spatial coherence. Available bandwidth on glass lasers is sufficient to suppress speckle self-focusing, which evolves on speckle width acoustic time scales, but cannot approach what is needed[38] to suppress SRS.

Acknowledgements: We are pleased to acknowledge insightful conversations with R. L. Berger, R. P. Drake and W. Seka. H. Rose was supported by the New Mexico Consortium and Department of Energy Award No. DE-SCOO02238.

[38] Philippe Mounaix, "Effects of laser spatial incoherence with finite correlation length on the space-time behavior of backscattering instabilities," *Physical Review E* 52, no. 2 (1995): R1306.